\documentclass[prd,aps,10pt,showkeys,nofootinbib,showpacs,twocolumn]{revtex4-2}

\usepackage{amsmath,amssymb,amsfonts,amsthm}
\usepackage{amsbsy} 
\usepackage{epsfig}
\usepackage{latexsym}
\input amssym.def
\input amssym.tex

\usepackage{hyperref}
\usepackage{xcolor}

\newcommand{\oarX}[1]{\href{http://arxiv.org/abs/#1}{{\ttfamily #1}}}
\newcommand{\arX}[1]{\href{http://arxiv.org/abs/#1}{{\ttfamily arXiv:#1}}}
\newcommand{\doin}[2]{\href{http://dx.doi.org/#1}{#2}}

\graphicspath{{figures/}} 

\def\barr{\begin{array}}
\def\earr{\end{array}}

\def\ben{\begin{equation}}
\def\een{\end{equation}}
\def\bs{\begin{subequations}}
\def\es{\end{subequations}}
\def\bena{\begin{eqnarray}}
\def\eena{\end{eqnarray}}

\def\im{{\rm i}}

\def\be{\begin{equation}}
\def\ee{\end{equation}}

\def\bes{\begin{eqnarray}}
\def\ees{\end{eqnarray}}

\newcommand{\dd}{\mathrm{d}}

\newcommand{\braket}[1]{\langle #1 \rangle}
\newcommand{\bra}[1]{\left\langle #1 \right|}
\newcommand{\ket}[1]{\left| #1 \right\rangle}

\begin{document}

\title{Stationary cosmology in group field theory}

\author{Steffen Gielen}
\affiliation{School of Mathematics and Statistics, University of Sheffield, Hicks Building, Hounsfield Road, Sheffield S3 7RH, United Kingdom}
\email{s.c.gielen@sheffield.ac.uk}
\author{Robert Santacruz}
\affiliation{Department of Mathematics and Statistics, University of New Brunswick, Fredericton, New Brunswick, Canada E3B 5A3}
\email{robert.santacruz@unb.ca}
\affiliation{School of Mathematics and Statistics, University of Sheffield,  Hicks Building, Hounsfield Road, Sheffield S3 7RH, United Kingdom.}
\date{\today}


\begin{abstract}
Group field theory (GFT) models for quantum gravity coupled to a massless scalar field give rise to cosmological models that reproduce the (expanding or contracting) dynamics of homogeneous and isotropic spacetimes in general relativity at low energies, while including high-energy corrections that lead to singularity resolution by a ``bounce.'' Here we investigate two possibilities for obtaining stationary solutions in GFT cosmology, which could be useful as an analogue of Minkowski spacetime. We first focus on a limit in which interactions are neglected and the effective Newton's constant in GFT cosmology is taken to zero. In this limit, we derive an effective Friedmann equation that shows no stationary solutions, but departures from the trivial classical dynamics falling off rapidly, similar to the usual correction terms responsible for the bounce. Since the effective Newton's constant needs to be exactly zero, the scenario is fine-tuned. A more satisfactory approach is obtained in a weakly interacting model: we find bound states with sharply peaked volume, representing a stationary semiclassical cosmology, and show that coherent states peaked around the minimum of the potential remain stable with small quantum fluctuations, and only small oscillations around a nearly constant volume. These coherent states realise the idea of a ``quantum gravity condensate.''
\end{abstract}


\maketitle

\section{Introduction}

Many approaches to quantum gravity entertain the idea that space and time are not fundamental structures that all of physics is built on, but themselves ``emergent'' from other quantum or discrete degrees of freedom with no initial spacetime continuum \cite{DanielePhilo}. A fundamental challenge is then to show how the usual classical, continuum nature of space and time might be recovered, at least in an approximation or perhaps in one out of different possible phases of a statistical description (see, e.g., Ref.~\cite{CDTreview} for the example of causal dynamical triangulations). One might look at other examples of emergence in physics such as a macroscopic electromagnetic field defined as a coherent state in quantum electrodynamics, or an effective continuum superfluid description of Bose--Einstein condensates in a quantum field theory of atoms. The latter in particular has served as an inspiration for looking for spacetime as a kind of Bose--Einstein condensate of quantum gravity ``atoms,'' i.e., a nonperturbative ground state away from the usual (Fock) vacuum \cite{QGcondensate}. In the group field theory (GFT) approach to quantum gravity \cite{GFTreview}, following this approach seems rather natural since the fundamental degrees of freedom of the ``group field'' are directly interpreted as quanta of spacetime, or elementary building blocks of spin networks in the language of loop quantum gravity \cite{LQGGFT}. The initial (perturbative) GFT vacuum contains no quanta, and hence no spacetime, as is manifest by the fact that quantities like areas or volumes vanish; a macroscopic geometry must be built up from many excitations over this initial vacuum.

The idea that continuum spacetime could emerge from a phase transition to GFT condensate was proposed in earlier papers \cite{DanieleCondensate} and then implemented concretely by building on a particular prescription for canonical quantisation \cite{GFTcondearly}. One basic question in an emergent spacetime scenario is how to define dynamics in a system without any fundamental notion of time. Here one can follow ideas from canonical quantum gravity and quantum cosmology \cite{MatterClocksQG} and introduce matter degrees of freedom that can play the role of a (relational) clock. Following this idea and coupling a massless ``clock'' scalar field to gravity in GFT, emergent Friedmann equations for the relational volume (i.e., for a volume of space given as a function of the scalar field) can be derived, showing agreement with general relativity at large volumes and singularity resolution by a bounce \cite{CosmoGFT}. The resulting cosmology resembles very closely that of loop quantum cosmology (LQC), raising the hope that GFT could provide an embedding of LQC into full quantum gravity.

The results of Ref.~\cite{CosmoGFT} were obtained using a number of simplifying assumptions; in particular, one usually restricts to a single mode in the expansion of the group field into Peter--Weyl modes, and interactions are (initially) neglected. One also works in a mean-field approximation, assuming a type of coherent state. The last approximation is inspired by the idea of a quantum gravity condensate and by the requirement that any cosmology emergent from quantum gravity should be semiclassical, with small fluctuations over expectation values for geometric observables. However, if interactions are neglected, what leads to the emergence of a macroscopic geometry is not so much a process of condensation but rather an instability in the free (linear) theory: in this approximation the dynamics of a single field mode resembles that of an upside-down harmonic oscillator, whose classical solutions grow or decay exponentially, just as the volume of the corresponding classical cosmology. This exponential behaviour of solutions was studied more explicitly, e.g., in Ref.~\cite{lowspin}, and a more general analysis of the quantum theory in a deparametrised approach was given in Ref.~\cite{EdHamilt}. Here, by choosing the scalar matter field as a clock before quantisation, one obtains a standard Hamiltonian acting on a Fock space generated from creation and annihilation operators associated to the upside-down harmonic oscillator. The Hamiltonian is quadratic in these, and corresponds to a squeezing operator (realising the proposal of Ref.~\cite{CosmoSqueezing}). The Fock ``vacuum'' defined by $\hat{a}_J|0\rangle=0$ for a mode $J$ is unstable and the number of quanta with respect to it grows exponentially under time evolution. It is then not surprising that almost {\em any} quantum state in this truncation leads to an effective Friedmann equation for the expectation value of the volume that reduces to that of general relativity at low energies and includes a bounce \cite{AxelGeneralCosmo}. The requirement that the state be semiclassical at late times is nontrivial and still suggests that one should work, e.g., with Fock coherent states.

To extend the results of Ref.~\cite{CosmoGFT} beyond the approximation of negligible interactions, the effect of certain interaction terms was included into the derivation of effective Friedmann equations in Ref.~\cite{AndreasMairiCosmo}. Other assumptions, in particular the mean-field approximation, were maintained. One finds that any monomial interaction term in GFT can be mapped to an additional term in the effective Friedmann equation, analogous to a perfect fluid contribution whose equation of state is related to the field power in the interaction. In this way, effective contributions corresponding to dust, dark energy, or other matter may in principle be obtained. However, given that these new terms become relevant when interactions are strong, one expects the mean-field approximation to break down, as explained in Ref.~\cite{CosmoGFT} and shown explicitly in Ref.~\cite{AxelGeneralCosmo}.

The recovery of expanding solutions that mimic the dynamics of classical general relativity coupled to a massless scalar field is an important result, but one might be interested in stationary solutions as well. Given that the expansion in usual GFT cosmology is driven by the energy density in the scalar field, is there a way to switch it off? Here we consider two approaches towards addressing this question. The first corresponds to the idea of taking a ``$G\rightarrow 0$'' limit in GFT cosmology, as is sometimes considered in other approaches to quantum gravity \cite{Gtozero}. Newton's constant $G$ appears emergent from a combination of fundamental couplings in the GFT action \cite{CosmoGFT,lowspin}, so this is the limit of a vanishing coupling in the GFT action. We find that this procedure does indeed modify the late-time behaviour of GFT cosmology in the expected way, leading to an asymptotically stationary geometry. However, there are still high-energy corrections similar to the ones causing the bounce in usual GFT cosmology, so that we do not obtain a truly stationary solution. These results are perhaps expected, but involve some interesting subtleties. In particular, we now need to introduce creation and annihilation operators for a system analogous to a free particle in quantum mechanics, which requires using an arbitrary length scale (see, e.g., Ref.~\cite{FreeParticleCoh}). This length scale enters geometric observables in GFT, whose meaning is hence ambiguous. The assumption of an exactly vanishing coupling also constitutes fine-tuning. 

A different approach is to include interactions and look for dynamically stationary solutions. Our approach follows the one of Ref.~\cite{AndreasMairiCosmo} without relying on a mean-field approximation: we are looking for exact solutions of the interacting theory. Finding such solutions requires numerical methods, but can be done to arbitrary precision. We can identify bound states in which all expectation values remain constant, so that they might be seen as representing a stationary cosmology. While these states are not peaked on a particular value of the group field, they have small fluctuations in the cosmologically more relevant volume (or number of quanta) and can hence be seen as semiclassical. We also turn to the familiar proposal of coherent states, peaked around the minimum of the potential. We show that these states are stable with small quantum fluctuations, even though they are not exactly stationary and show small oscillations in quantities like the volume. Both the exact bound state solutions and the coherent states we have constructed are promising candidates for an emergent semiclassical and stationary or almost stationary spacetime; they represent a cosmology in which the contribution to the effective energy density coming from GFT interactions cancels the terms usually responsible for expansion. This proposal could be argued to be the most explicit realisation of a ``quantum gravity condensate'' achieved so far, albeit in a relatively simple toy model. 

\section{Brief overview of GFT cosmology}

Here we review the derivation of effective cosmological dynamics from GFT in the deparametrised approach of Ref.~\cite{EdHamilt}. Although this formalism differs in its assumptions and motivations from the ``algebraic'' canonical quantisation first proposed in Refs.~\cite{LQGGFT,GFTcondearly}, at the level of effective cosmology the two approaches lead to rather similar results; see, e.g., Ref.~\cite{AndreaGFT} for a recent review comparing the two. What is important to obtain a particular cosmology is whether interactions or multiple field modes are included, as well as whether one assumes a coherent state. We will restrict to a single field mode throughout.

For GFT models for quantum gravity coupled to a massless scalar field, a common starting point is a (real) field $\varphi$ whose arguments are four ${\rm SU}(2)$ group elements, corresponding to parallel transport variables of discrete gravity in the Ashtekar--Barbero formalism, and a real-valued argument $\chi$ corresponding to the matter scalar field. The field is usually assumed to satisfy the ``gauge invariance'' property
\be
\varphi(g_1,\ldots,g_4,\chi)=\varphi(g_1h,\ldots,g_4h,\chi)
\label{gaugeinv}
\ee
for any $h\in{\rm SU}(2)$. If we picture the elementary excitations of this quantum field as spin-network vertices (labelled by $\chi$) with four open links labelled by the $g_I$, this property ensures that GFT states are invariant with respect to discrete ${\rm SU}(2)$ gauge transformations acting on these vertices. 

Assuming that $\varphi$ is square integrable on ${\rm SU}(2)^4$, we can define a Peter--Weyl decomposition as
\be
\varphi(g_I,\chi) = \sum_{j_I,m_I,n_I,\iota}\varphi^{j_I,\iota}_{m_I}(\chi)\mathcal{I}^{j_I,\iota}_{n_I}\prod_{a=1}^4 \sqrt{2j_a+1}\,D^{j_a}_{m_a\,n_a}(g_a)
\label{peterweyl}
\ee
where $j_I\in\{0,\frac{1}{2},1\ldots\}$ are irreducible representations of ${\rm SU}(2)$, $m_I$ and $n_I$ are magnetic indices taking values between $-j_I$ and $j_I$, and $\mathcal{I}$ are a basis of intertwiners (indexed by $\iota$) compatible with the chosen $j_I$, which are needed in order to satisfy Eq.~(\ref{gaugeinv}). $D^j_{m\,n}(g)$ are the Wigner $D$-matrices in the representation $j$. It is very convenient to introduce a multi-index $J\equiv(j_I,m_I,\iota)$ so that the field modes in Eq.~(\ref{peterweyl}) become more simply $\varphi_J(\chi)$.

Following Ref.~\cite{EdHamilt} we will assume an action
\begin{align}
S&=\,\frac{1}{2}\int {\rm d}^4 g\;{\rm d}\chi\;\varphi(g_I,\chi)\left(K^{(0)}+K^{(2)}\partial_\chi^2\right)\varphi(g_I,\chi)-V[\varphi]\nonumber
\\&=\,\frac{1}{2}\sum_J\int{\rm d}\chi\;\varphi_{-J}(\chi)\left(\mathcal{K}^{(0)}_J+\mathcal{K}^{(2)}_J\partial_\chi^2\right)\varphi_J(\chi)-V[\varphi]\,,\label{action}
\end{align}
where in the second line we have used the Peter--Weyl decomposition and $-J\equiv(j_I,-m_I,\iota)$ denotes flipping of the magnetic indices (needed to ensure a real Lagrangian). $K^{(0)}$ and $K^{(2)}$ can contain derivative operators with respect to the ${\rm SU}(2)$ variables, in particular Laplace--Beltrami operators, which become diagonal in the second line, so that $\mathcal{K}^{(0)}_J$ and $\mathcal{K}^{(2)}_J$ are just $J$-dependent numbers. More generally, higher order derivatives in $\chi$ could be present, but one can see Eq.~(\ref{action}) as a truncation in derivatives (as proposed in Ref.~\cite{CosmoGFT}) or as a definition of the fundamental theory. A first derivative term is forbidden by the symmetry of the action under $\chi\rightarrow-\chi$ which is required as a symmetry of relativistic matter fields. $V[\varphi]$ includes all interactions, i.e., terms higher than second order in $\varphi$, whose structure is model-dependent. 

One can now proceed with canonical quantisation based on promoting $\varphi_J$ and its conjugate momentum
\be
\pi_J:=\frac{\partial\mathcal{L}}{\partial(\partial_\chi\varphi_J)}=-\mathcal{K}^{(2)}_J\partial_\chi\varphi_J
\ee
to operators satisfying the usual $[\hat\varphi_J(\chi),\hat\pi_{J'}(\chi)]={\rm i}\delta_{J,J'}$. In other words, the scalar field variable $\chi$ is now treated as a conventional time variable. The quadratic part of the Hamiltonian is a sum of single-mode Hamiltonians,
\begin{align}
\hat{\mathcal{H}}&=\,-\frac{1}{2}\sum_J\left[\frac{\hat\pi_J\hat\pi_{-J}}{\mathcal{K}^{(2)}_J}+\mathcal{K}^{(0)}_J\hat\varphi_J\hat\varphi_{-J}\right]+V[\hat\varphi]\nonumber
\\&=:\,\sum_J \hat{\mathcal{H}}_J + V[\hat\varphi]\,.
\label{Hamiltonian}
\end{align}
For modes for which $\mathcal{K}^{(0)}_J$ and $\mathcal{K}^{(2)}_J$  have the same sign, this quadratic part is the Hamiltonian of a harmonic oscillator (potentially with an unusual minus sign), whereas for opposite signs the Hamiltonian is that of an upside-down harmonic oscillator with negative quadratic potential. It is the second case which is relevant for cosmology since it has exponentially growing solutions, and most of the literature is focused on this case.

Introducing for each $J$ an annihilation operator
\be
\hat{a}_J = \frac{1}{\sqrt{2\omega_J}}\left(\omega_J\hat\varphi_J + {\rm i}\,\hat\pi_J^\dagger\right)
\label{annihilation}
\ee
and its conjugate (creation operator) $\hat{a}^\dagger_J$, where $\omega_J:=\sqrt{|\mathcal{K}^{(0)}_J \mathcal{K}^{(2)}_J|}$, the quadratic Hamiltonian for one of the unstable modes becomes
\be
\hat{\mathcal{H}}_J = \frac{1}{2}M_J\left(\hat{a}^\dagger_J \hat{a}^\dagger_{-J} + \hat{a}_J \hat{a}_{-J}\right)
\label{squeezingham}
\ee
with $M_J:=-{\rm sgn}\left(\mathcal{K}^{(0)}_J\right)\sqrt{|\mathcal{K}^{(0)}_J/\mathcal{K}^{(2)}_J|}$ (see Ref.~\cite{EdHamilt} for details). If we neglect the effect of interactions contained in $V[\hat\varphi]$ for now, we see that this Hamiltonian takes the form of a squeezing operator, in that time evolution with respect to $\hat{\mathcal{H}}_J$ transforms the vacuum into a squeezed state (or more general initial states into generalised squeezed states). If one works in the Heisenberg picture, the operators $\hat{a}_J$ and $\hat{a}^\dagger_J$ are time-dependent with
\begin{align}
\hat{a}_J(\chi)&=\,\hat{a}_J(0)\cosh(M_J\chi)-\im\,\hat{a}^\dagger_{-J}(0)\sinh(M_J\chi)\,,\nonumber
\\\hat{a}^\dagger_J(\chi)&=\,\hat{a}^\dagger_J(0)\cosh(M_J\chi)+\im\,\hat{a}_{-J}(0)\sinh(M_J\chi)\,.
\end{align}
Likewise, the number operator $\hat{N}_J:=\hat{a}^\dagger_J \hat{a}_J$ can be written as 
\begin{align}
\hat{N}_J(\chi)&=\,\frac{1}{2}\left(\hat{N}_J(0)+\hat{N}_{-J}(0)+1\right)\cosh(2M_J\chi)\nonumber
\\&\;\;+\frac{1}{2}\left(\hat{N}_J(0)-\hat{N}_{-J}(0)-1\right)
\label{numberopevolve}
\\&\;\;+\frac{\im}{2}\left(\hat{a}_J(0)\hat{a}_{-J}(0)-\hat{a}^\dagger_J(0)\hat{a}^\dagger_{-J}(0)\right)\sinh(2M_J\chi)\nonumber
\end{align}
showing explicitly that the particle number grows exponentially in $\chi$ for arbitrary initial states. The quanta generated by actions of $\hat{a}^\dagger_J$ are interpreted as ``atoms'' of geometry in the sense of loop quantum gravity \cite{LQGGFT}, which are assigned a volume $V_J$ dependent on the representation labels $J$. If one assumes that only a single field mode is excited, the total volume is simply proportional to the number of quanta, $\langle\hat{V}\rangle= V_J\langle\hat{N}_J\rangle$. This assumption is most easily made self-consistent by focusing on a mode for which all magnetic indices vanish, so that $J=-J$; Eq.~(\ref{numberopevolve}) then refers to operators for a single mode only.

For this case, one can easily show that $V(\chi):=\langle\hat{V}(\chi)\rangle$ satisfies the differential equation \cite{AxelGeneralCosmo}
\be
\left(\frac{V'(\chi)}{V(\chi)}\right)^2=4M_J^2\left(1+\frac{V_J}{V(\chi)}+\frac{K_0^2-V_J\,V(0)-V(0)^2}{V(\chi)^2}\right)
\label{friedmann}
\ee
with $K_0:=\frac{\im}{2}V_J\left(\langle\hat{a}_J(0)\hat{a}_{-J}(0)\rangle-\langle\hat{a}^\dagger_J(0)\hat{a}^\dagger_{-J}(0)\rangle\right)$. This is the analogue of the Friedmann equation, and can be used to interpret the expectation value $V(\chi)$ of the volume in cosmological terms.

Comparing Eq.~(\ref{friedmann}) with its general relativity analogue $(V'/V)^2=12\pi G$, the first observation is that Eq.~(\ref{friedmann}) reduces to general relativity at large volume, provided that $M_J^2=3\pi G$ where $G$ is Newton's constant. In this sense, we can say that Newton's constant is emergent from fundamental GFT couplings. At smaller volumes, there are corrections to general relativity, in particular a $1/V^2$ term which is almost always repulsive (there are fine-tuned initial conditions for which $K_0^2-V_J\,V(0)-V(0)^2$ can vanish, otherwise its sign is negative). When it is repulsive, it will dominate at small volume, leading to a bounce that resolves the classical singularity; in other words, $V(\chi)$ never reaches zero. 

These conclusions do not depend on a choice of state; however, for a semiclassical interpretation one should also require that fluctuations in quantities like the volume are small at late times, $\Delta V \ll \langle\hat{V}\rangle$, which suggests that one should choose, e.g., a Fock coherent state satisfying (again in the Heisenberg picture)  \cite{AxelGeneralCosmo}
\be
\hat{a}_J(0)|\alpha\rangle = \alpha|\alpha\rangle\,.
\label{coherentdef}
\ee
If either multiple field modes or interactions in $V[\hat\varphi]$ are included, the situation is more complicated; in the first case and without interactions, one still has Eq.~(\ref{numberopevolve}) for each mode and while deriving an equation for $(V'/V)^2$ is still straightforward, the right-hand side is complicated and does not admit a simple cosmological interpretation as in Eq.~(\ref{friedmann}). Interactions would generally couple different modes and spoil the property of independent evolution. As a first step, one can study toy models with a single self-interacting mode as done, e.g., in Refs.~\cite{AxelGeneralCosmo,AndreasMairiCosmo}; we will study such a toy model below. In this case, one deals with the quantum theory of an upside-down harmonic oscillator with a higher-order potential, for which there are generally no analytic solutions. One can still propose a mean-field approximation to solve essentially classical equations as in Ref.~\cite{AndreasMairiCosmo}, although such an approximation will break down once interactions become important. Numerical studies as in Ref.~\cite{AxelGeneralCosmo} are an alternative possibility.

\section{Vanishing Newton's constant}

An interesting question which has so far escaped detailed attention is what happens in the case that $M_J$ vanishes, i.e., the case where $\mathcal{K}^{(0)}_J$ is zero for a particular mode $J$. Given that the indices contained in $J$ are discrete, there is no particular reason to expect that such a $J$ exists, but one might assume that it does. In this case, while the Legendre transform leading to a Hamiltonian (\ref{Hamiltonian}) can be defined as before, the creation and annihilation operators used above become ill-defined since $\omega_J\rightarrow 0$. Indeed, one now faces the problem of defining creation and annihilation operators for a system equivalent to a free particle in quantum mechanics, rather than a (regular or upside-down) harmonic oscillator.

For simplicity, we restrict to a single mode with $J=-J$, and assume that $\mathcal{K}^{(0)}_J$ vanishes. We will, for now, also neglect interactions. With all these approximations we obtain a quadratic Hamiltonian
\be
\hat{\mathcal{H}}_J = -\frac{1}{2}\frac{\hat\pi_J^2}{\mathcal{K}^{(2)}_J}
\ee
which is the Hamiltonian of a free particle in one dimension whose quantum theory is, of course, well known. However, here we are interested in the interpretation of the corresponding GFT cosmology, which requires defining a number operator $\hat{N}_J=\hat{a}^\dagger_J \hat{a}_J$ in terms of some suitable ladder operators $\hat{a}_J$ and $\hat{a}_J^\dagger$. The definition (\ref{annihilation}) cannot be applied in this case, but one can define
\be
\hat{a}_J = \frac{1}{\sqrt{2\omega_0}}\left(\omega_0\hat\varphi_J + {\rm i}\,\hat\pi_J\right)
\ee
where $\omega_0$ is now an arbitrary scale rather than derived from the Hamiltonian. We then have
\be
\hat{\mathcal{H}}_J = \frac{\omega_0}{4\mathcal{K}^{(2)}_J}(\hat{a}^\dagger_J-\hat{a}_J)^2 
\ee
which decomposes into the difference of a squeezing operator similar to Eq.~(\ref{squeezingham}) and a standard harmonic oscillator Hamiltonian $\propto(\hat{a}^\dagger_J\hat{a}_J+\hat{a}_J\hat{a}^\dagger_J)$, i.e., the difference of an operator with continuous and one with discrete spectrum. The overall spectrum is of course continuous, but the number operator $\hat{a}^\dagger_J \hat{a}_J$ has the usual spectrum given by the non-negative integers, since that simply derives from the algebraic relation $[\hat{a}_J,\hat{a}^\dagger_J]=1$. We can then go ahead and define an effective volume operator $\hat{V}=V_J\hat{N}_J$ as in usual GFT cosmology.

The Heisenberg equations of motion are now
\be
\frac{\dd\hat{a}_J}{\dd\chi}=-\im \frac{\omega_0}{2\mathcal{K}^{(2)}_J}\left(\hat{a}^\dagger_J-\hat{a}_J\right)
\ee
and its Hermitian conjugate, with solution
\be
\hat{a}_J(\chi)=\hat{a}_J(0)-\sqrt{\frac{\omega_0}{2}}\frac{1}{\mathcal{K}^{(2)}_J}\hat{\pi}\,\chi
\ee
and Hermitian conjugate; $\hat\pi$ is time-independent since it commutes with the Hamiltonian. This solution of course represents the linear relation between ``position'' and ``time'' expected for the free particle.

For the number operator $\hat{N}_J=\hat{a}^\dagger_J \hat{a}_J$ we then find
\begin{align}
\hat{N}_J(\chi)&=\hat{N}_J(0)-\sqrt{\frac{\omega_0}{2}}\frac{1}{\mathcal{K}^{(2)}_J}\left(\hat{a}^\dagger_J(0)\hat{\pi}+\hat{\pi}\hat{a}_J(0)\right)\chi\nonumber
\\&\quad-\frac{\omega_0}{\mathcal{K}^{(2)}_J}\hat{\mathcal{H}}_J\,\chi^2
\label{nofchi}
\end{align}
and hence quadratic growth in the volume with respect to $\chi$. Since there are no states of zero energy (a putative eigenstate of zero momentum would not be normalisable), this general behaviour applies to all states and there no exactly stationary solutions. On the other hand, one can derive an effective Friedmann equation
\begin{align}
\left(\frac{V'(\chi)}{V(\chi)}\right)^2 =&\, -\frac{4\omega_0\,E}{\mathcal{K}^{(2)}_J}\,\frac{V_J}{V(\chi)}+\frac{\omega_0}{\mathcal{K}^{(2)}_J}A\frac{V_J^2}{V(\chi)^2}\,,
\label{zeroGfriedmann}
\\A=&\,\frac{C_0^2}{2\mathcal{K}^{(2)}_J}+4N_0E
\end{align}
where $V(\chi)=\langle \hat{V}(\chi)\rangle$ as before, $E=\langle\hat{\mathcal{H}}_J\rangle$ is the expectation value of the Hamiltonian, $N_0=\langle \hat{N}_J(0)\rangle$ is the average initial particle number, and $C_0=\langle \hat{a}^\dagger_J(0)\hat{\pi}+\hat{\pi}\hat{a}_J(0)\rangle$.
Since the inequality $\langle\hat{N}_J(\chi)\rangle\ge 0$ for all $\chi$ implies $A\le 0$, the $1/V^2$ term in the effective Friedmann equation is repulsive for small volumes and generically (for $A<0$) guarantees that the volume never reaches zero.

At late (or very early) times when the volume is large, the right-hand side of Eq.~(\ref{zeroGfriedmann}) goes to zero and the emergent spacetime geometry becomes approximately flat: the terms on the right-hand side of Eq.~(\ref{zeroGfriedmann}) are of the same form as the subleading corrections in Eq.~(\ref{friedmann}). In both cases, these can be seen as quantum gravity corrections to the correct classical limit. In this sense, the general structure of Eq.~(\ref{zeroGfriedmann}) might be expected: while the emergent Newton's constant could be fine-tuned to zero, there is not a single limit in the quantum gravity framework of GFT that would also make all the subleading corrections vanish. These subleading corrections are suppressed by inverse powers in the number of GFT quanta, which we expect to be large for a semiclassical interpretation.

From Eq.~(\ref{nofchi}) we see that at late (or very early) times, the relative uncertainty in the volume asymptotes to
\be
\label{eq:asymprel}
\frac{(\Delta_V)^2}{V(\chi)^2} = \frac{\langle \hat{V}^2(\chi)\rangle - V^2}{V^2} \rightarrow \frac{(\Delta_H)^2}{E^2}
\ee
(where we use the notation $(\Delta_O)^2:=\langle\hat{O}^2\rangle-\langle\hat{O}\rangle^2$), which can be made arbitrarily small by choosing states sharply peaked around an average energy value $E$. Hence there exists a large class of states that evolve into semiclassical, asymptotically flat effective geometries. 

This notion of semiclassicality, based on relative uncertainty in the volume, does not mean that states remain sharply peaked in quantities such as the field $\varphi_J$ or momentum $\pi_J$. For instance, if we define coherent states as proposed in  Ref.~\cite{FreeParticleCoh}, we can see that uncertainties grow as we move away from the initial time $\chi=0$,
\begin{align}
	(\Delta_{\varphi_J})^2 &= \frac{1}{2}\left(\frac{1}{\omega_0}+\frac{\omega_0}{\mathcal{K}_J^{(2)}}\,\chi^2\right)\,,\\
	(\Delta_{\pi_J})^2 &= \frac{\omega_0}{2}\,.
\end{align}
From these expressions we can readily see that Fock coherent states do not stay coherent, as $\Delta_{\varphi_J}\Delta_{\pi_J}=\frac{1}{2}$ only at the initial time. This behaviour seems to be general for Hamiltonians that do not commute with $\hat{N}_J$. Due to Eq.~\eqref{eq:asymprel}, such states can still be made sharply peaked around a given volume for early and late times.

Given the use of $\chi$ as a clock, the energy $E$ is usually interpreted as representing the momentum conjugate to the scalar matter field \cite{AxelGeneralCosmo}. It seems puzzling that in this model the energy is restricted to be negative, so that this momentum would have a preferred sign in contrast with classical cosmology, where it is simply related to the time derivative in the scalar field which can take either sign. Moreover, Eq.~(\ref{zeroGfriedmann}) also depends explicitly on the arbitrary scale $\omega_0$, since the number operator itself required this scale for its definition. In this sense, the meaning of GFT geometric observables in this scenario seems ambiguous, so that it would seem difficult to extract any phenomenology from it. This is in contrast to the usual case Eq.~(\ref{friedmann}) which involves no additional arbitrary scales.

Perhaps the most unphysical aspect of this scenario is the fine-tuning in setting $\mathcal{K}^{(0)}_J$ to zero. As we mentioned, there will generically be no $J$ which satisfies this property; even if there is such a $J$, $\mathcal{K}^{(0)}_J$ will be non-zero for other modes and there will generally still be modes satisfying Eq.~(\ref{numberopevolve}) and growing exponentially. The model has to be set up in a specific way for no such modes to exist, and would be unstable under inclusion of other modes.

\section{Interacting GFT model}

To address some of the issues with the GFT cosmology scenario obtained from tuning $\mathcal{K}^{(0)}_J$ to zero, we turn to a second approach, in which the quadratic Hamiltonian is unchanged, but one now includes interaction terms as well. The idea is that the exponential instability seen in Eq.~(\ref{numberopevolve}), which arises from a quadratic Hamiltonian unbounded from below, is an artefact of neglecting interactions; the full theory should have a Hamiltonian that is bounded from below. This viewpoint was advocated in Ref.~\cite{AndreasMairiCosmo}, in the context of a mean-field approximation, and used to derive an effective GFT cosmology for a simple interacting toy model. Here we will present numerical evolution of the quantum theory, which can help to understand the validity of the mean-field approximation.

As before, we restrict the analysis to a single Peter--Weyl mode with $J=-J$. We then add a $\varphi^4$ interaction term to the Hamiltonian (\ref{squeezingham}) to obtain
\be
\hat{\mathcal{H}}_J = \frac{1}{2}M_J\left(\hat{a}^\dagger_J \hat{a}^\dagger_{J} + \hat{a}_J \hat{a}_{J}\right) + \frac{g}{4} |M_J|\left(\hat{a}_J+\hat{a}_J^\dagger\right)^4
\label{interactHam}
\ee
where $0<g\ll 1$, and we now assume $\mathcal{K}^{(0)}_J>0$ and $\mathcal{K}^{(2)}_J<0$ (the opposite sign choice can be treated analogously). In most GFT models for quantum gravity, interactions couple different modes, e.g., to encode matching conditions expected from gluing tetrahedra to higher-dimensional structures \cite{GFTreview}. We take this ``local'' interaction in $J$ as a general toy model for quantum behaviour of the GFT field, keeping in mind that choosing particularly symmetric GFT states can reduce more general interactions to local ones \cite{CosmoGFT}.

\begin{figure}[h]
\centering
\includegraphics[width=\columnwidth]{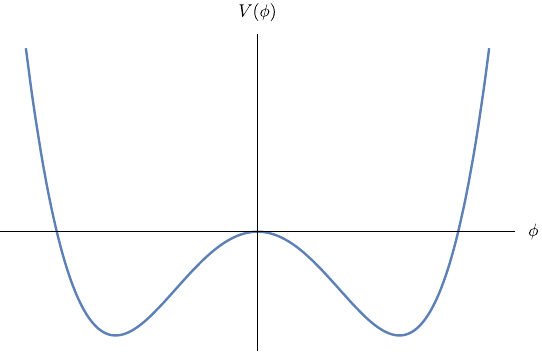}
\caption{Schematic plot of the potential for our GFT toy model Hamiltonian.}
\label{potentialplot}
\end{figure}

The previous interacting Hamiltonian is equivalent to a quantum mechanical system in terms of $\hat\varphi_J$ and $\hat\pi_J$,
\begin{align}
\hat{\mathcal{H}}_J &=\, -\frac{1}{2}\left[\frac{\hat\pi_J^2}{\mathcal{K}^{(2)}_J}+\mathcal{K}^{(0)}_J\hat\varphi_J^2\right]+ \frac{\tilde{g}}{4} \,\hat\varphi_J^4\,,
\\\tilde{g}&=\,4g \,\sqrt{\left|\left(\mathcal{K}^{(0)}_J\right)^3\mathcal{K}^{(2)}_J\right|}\,.
\end{align}
In a mean-field approximation, we would replace $\hat\pi_J$ and $\hat\varphi_J$ by their respective expectation values $p_J$ and $\phi_J$. We then obtain an effectively classical Hamiltonian
\be
\mathcal{H} = -\frac{1}{2}\left(\frac{p_J^2}{\mathcal{K}^{(2)}_J}+\mathcal{K}^{(0)}_J\phi_J^2\right)+\frac{\tilde{g}}{4}\,\phi_J^4\,.
\label{classHamilt}
\ee
Stationary solutions of the resulting equations of motion correspond to extrema of this Hamiltonian in $\phi_J$, for $p_J=0$, given by $\phi_J=0$ and
\be
\phi_J = \phi_J^{(\pm)} = \pm \sqrt{\frac{\mathcal{K}^{(0)}_J}{\tilde{g}}}\,.
\label{minima}
\ee 
This mean field model is equivalent to a classical system with a potential $U[\phi_J] = -\frac{\mathcal{K}^{(0)}_J}{2}\phi_J^2+\frac{\tilde{g}}{4}\phi_J^4$, usually referred to as double well potential (see Fig.~\ref{potentialplot}). The values of the field at the bottom of the potential imply a minimum value for the energy and volume
\begin{align}
\label{eq:classemin}
E_{{\rm min}} &= U\Big|_{\phi_J=\phi_J^{(\pm)}}=-\frac{|M_J|}{16g}\,,\\
\label{eq:classvolmin}
V_{{\rm min}} &= \frac{V_J\,\omega_J}{2}\phi_J^2\Big|_{\phi_J=\phi_J^{(\pm)}} = \frac{V_J}{8g}\,.
\end{align}

We can attempt an interpretation of this stabilising behaviour in terms of GFT cosmology, as was done in Ref.~\cite{AxelGeneralCosmo} using a classical analogue system in which one treats the independent quadratic combinations $\hat{a}_J^2$, $(\hat{a}^\dagger_J)^2$ and $\hat{a}_J^\dagger\hat{a}_J$ as classical variables, ignoring higher order corrections coming from commutators of such variables. In such an approximation, one can derive an effective Friedmann equation
\begin{widetext}
\begin{align}
\left(\frac{V'(\chi)}{V(\chi)}\right)^2 &=\, -\frac{2M_J^2\,V_J^2}{g^2V(\chi)^2}\left(1+4g\left(\frac{E}{|M_J|}-\frac{V(\chi)}{V_J}\right)\right)\times\left[1-4g\frac{V(\chi)}{V_J}-\left(1-2g\frac{V(\chi)}{V_J}\right)\sqrt{1+4g\left(\frac{E}{|M_J|}-\frac{V(\chi)}{V_J}\right)}\right.\nonumber
\\&\quad\left.+2g\left(-\frac{3}{4}g+\frac{E}{|M_J|}\right)\right]
\label{generalFriedmann}
\end{align}
\end{widetext}
where $V(\chi)$ and $E$ are now effectively classical quantities, derived from the respective combinations of the fundamental variables. While this approximation differs from a simpler mean-field approximation, both can be seen as neglecting quantum corrections beyond a certain order.

We can substitute the classical minimum values \eqref{eq:classemin} and \eqref{eq:classvolmin} into Eq.~(\ref{generalFriedmann}) resulting in $(V'/V)^2=48g^2M_J^2$, with the only contribution coming from the last term, which arises from  a non-vanishing Casimir in the $\mathfrak{su}(1,1)$ algebra spanned by the basic operators \cite{AxelGeneralCosmo}, and may be seen as a quantum correction to the stationary classical dynamics. In simpler truncations where the right-hand side of Eq.~(\ref{generalFriedmann}) is only linear in all interaction couplings, as in Ref.~\cite{AndreasMairiCosmo}, this higher-order term would not be visible and a classical minimum would automatically be interpreted as a stationary cosmology. In either case, the effective Friedmann equation contains terms of both signs, so that cancellations can lead to a stationary solution. This can be compared to a classical cosmology for which, in the Friedmann equation
\be
\frac{\dot{a}^2}{a^2}=\frac{8\pi G}{3}\rho + \frac{\Lambda}{3}\,,
\ee
one chooses the energy density $\rho$ at a given time to exactly balance the contribution of a negative $\Lambda$; such a cosmology would however be unstable under perturbations. In GFT, it seems we can obtain a stationary cosmology by balancing the usual matter energy density with new contributions that appear to have effectively negative energy density, similar to a negative $\Lambda$.

\subsection{Full quantum analysis}
Turning to the full quantum theory, we return to the Schr\"odinger picture and aim to find stationary (or almost stationary) solutions to the Schr\"odinger equation for the model. The most obvious candidates for such states are eigenstates of the Hamiltonian (\ref{interactHam}), but we will also follow the more traditional approach in GFT cosmology to identify coherent states with suitable initial conditions that can have a good semiclassical interpretation.

Since this Hamiltonian is quartic in the basic ladder operators, there are no good methods for analytically deriving its spectrum and eigenstates. However, one can work numerically by representing the ladder operators as infinite matrices written in the basis of eigenstates of the number operator \cite{easycomp},
\begin{align}
\hat{a}_J & = \left(\begin{matrix} 0 & 1 & 0 & 0 & 0 & \ldots \cr 0 & 0 & \sqrt{2} & 0 & 0 & \ldots \cr 0 & 0 & 0 & \sqrt{3} & 0 & \ldots \cr 0 & 0 & 0 & 0 & 2 & \ldots \cr \vdots & \vdots & \vdots & \vdots & \vdots & \ddots\\ \end{matrix}\right)\,,
\label{amatrix}
\\\hat{a}^\dagger_J & = \left(\begin{matrix} 0 & 0 & 0 & 0 & 0 & \ldots \cr 1 & 0 & 0 & 0 & 0 & \ldots \cr 0 & \sqrt{2} & 0 & 0 & 0 & \ldots \cr 0 & 0 & \sqrt{3} & 0 & 0 & \ldots \cr \vdots & \vdots & \vdots & \vdots & \vdots & \ddots\\ \end{matrix}\right)\,.
\label{admatrix}
\end{align}

We may represent the basis states as
\be
\ket{0} = \begin{pmatrix}
				1\\
				0\\
				0\\
				\vdots\\
				\end{pmatrix}\,,\quad
\ket{1} = \begin{pmatrix}
				0\\
				1\\
				0\\
				\vdots\\
				\end{pmatrix}\,,\quad
\ket{2} = \begin{pmatrix}
				0\\
				0\\
				1\\
				\vdots\\
				\end{pmatrix}\,,\quad\ldots
\label{eq:basis}
\ee
One can express an operator as a matrix by writing it in terms of the ladder ones, $\hat{a}_J$ and $\hat{a}^{\dagger}_J$, provided that a truncation is used. This truncation sets a finite dimension for matrices, determines the accuracy of the calculations and can be extended for a higher accuracy of numerics. By representing the Hamiltonian as a truncated matrix, we can find its eigenvalues and eigenstates, and determine the the dynamics by expressing the Schr\"odinger equation as a matrix differential equation.

For small coupling constant $g$, there are a large number of bound states, with the ground state (lowest energy) close to the classical minimum, $\braket{\mathcal{H}}\sim E_{\rm min}=-\frac{|M_J|}{16g}$. Nonetheless, the expectation value of the field $\hat\varphi_J=(\hat{a}_J+\hat{a}_J^\dagger)/\sqrt{2\omega_J}$ is not near either of classical minima (at the bottom of the double well) but instead it is close to zero, with large fluctuations. 

\begin{figure}[htbp]
\centering
\includegraphics[width=1\columnwidth]{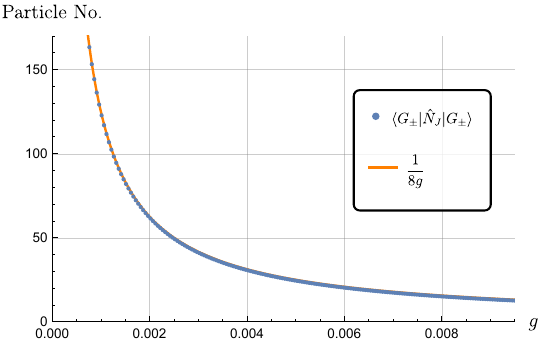}
\caption{Expectation value of the number of particles in the ground state as a function of the coupling constant $g$, compared with the classical result $1/8g$.}
\label{fig:nbound}
\end{figure}

\begin{figure}[htbp]
\centering
\includegraphics[width=1\columnwidth]{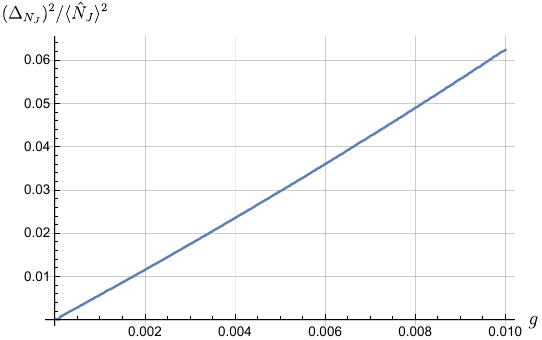}
\caption{Relative variance of the particle number $N_J$ for ground states $\ket{G_\pm}$. At low values of $g$, quantum fluctuations are still small. The relation is almost linear at small $g$, but becomes nonlinear as $g$ is increased further.}
\label{fig:nvariance}
\end{figure}

As an illustrative example, we fix the parameters to $\mathcal{K}^{(0)}_J=1,\;\mathcal{K}^{(2)}_J=-1,\;g=10^{-3}$; a matrix size of 500 gives good numerical accuracy. We then find a two-fold degenerate ground state $|G_\pm\rangle$ with
\begin{align}
\label{eq:Gpmexp1}
\langle G_\pm|\hat{\mathcal{H}}_J|G_\pm\rangle &\approx -61.79\,,
\\\langle G_\pm|\hat\varphi_J|G_\pm\rangle &\approx 0\,,
\\\langle G_\pm|(\hat\varphi_J)^2|G_\pm\rangle &\approx 249\,,
\\\bra{G_\pm}\hat{N}_J\ket{G_\pm} &\approx 124.5\,,
\\\bra{G_\pm}(\hat{N}_J)^2\ket{G_\pm} &\approx (124.85)^2\,,
\label{eq:Gpmexp2}
\end{align}
matching well with the classical $E_{\rm min}=-62.5$ but not with $\phi^{(0)}_J\approx 15.81$. We can see that, while such a state is not semiclassical in the group field $\varphi_J$, the state is sharply peaked around its expectation value of the volume $\langle\hat{V}\rangle= V_J\langle\hat{N}_J\rangle$. Somewhat surprisingly, such a bound state then already represents a semiclassical, stationary cosmology. In Fig.~\ref{fig:nbound} we show the expectation value $\bra{G_\pm}\hat{N}_J\ket{G_\pm}$ in the ground state(s) as a function of the coupling $g$; it follows closely the classical result given in Eq.~(\ref{eq:classvolmin}), $N_J=1/8g$. Fig.~\ref{fig:nvariance} shows that the relative variance $\frac{(\Delta_{N_J})^2}{\braket{\hat{N}_J}^2}$ monotonically increases with $g$, so that the semiclassical interpretation of the ground states breaks down at larger values of the coupling constant. For small $g$, this relative variance grows linearly in $g$ and hence scales as the inverse particle number; the relation becomes nonlinear at larger $g$. Moreover, the first higher energy states above the ground state show similar expectation values for the volume but with rapidly growing fluctuations, making those states less suitable for a semiclassical interpretation than the ground state.

Going beyond this simplest proposal, one can try to define some kind of coherent states from the eigenstates of this bounded Hamiltonian. These are called Gazeau--Klauder coherent states and have been studied for the double well-potential in Ref.~\cite{NovaesAguiar:2003}. Note then that, for an expectation value of the energy $\hat{\mathcal{H}}$ close to the minimum of the classical potential, this coherent state can be approximated by the two first eigenstates, and produces essentially the same expectation values as Eqs.~(\ref{eq:Gpmexp1})-(\ref{eq:Gpmexp2}).

Focusing on bound states is quite different from the traditional approach in GFT cosmology, in which semiclassical cosmology is represented via coherent Fock states \cite{GFTcondearly}. In our setting, we could define a (normalised) approximate coherent state at some initial time by
\be
|\alpha\rangle = e^{-|\alpha|^2/2}\sum_{i=0}^M \frac{\alpha^i}{\sqrt{i!}}|i\rangle
\ee
where $M$ is the cutoff on the truncation that one has applied to make the matrix representations (\ref{amatrix})-(\ref{admatrix}) finite. $M$ must be chosen large enough so that $\langle\alpha|\alpha\rangle=1$ up a small error below the accuracy one wants to work at. 

If we are interested in a state that represents the classically stationary configuration at the minimum of the potential, we can choose $\alpha$ such that the field is set at one of the classical minima, $\alpha=\pm\sqrt{1/8g}$. The value of $M$ then depends on $g$; for $g\sim 10^{-3}$ a value between 600 and 1000 is sufficient for very small errors (depending on the time of evolution).

\begin{figure}
\centering
\includegraphics[width=1\columnwidth]{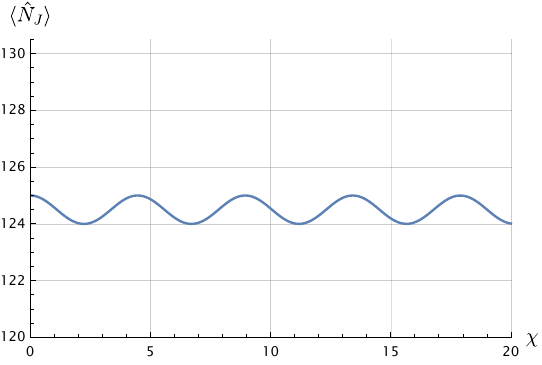}
\caption{Oscillatory evolution in relational time $\chi$ of the particle number expectation value in the coherent state representing the classically stationary state, for the set of parameters $\mathcal{K}^{(0)}_J = -1$, $\mathcal{K}^{(2)}_J = 1$ and $g=10^{-3}$.}
\label{fig:Nexp0}
\end{figure}

\begin{figure}[htbp]
\centering
\includegraphics[width=1\columnwidth]{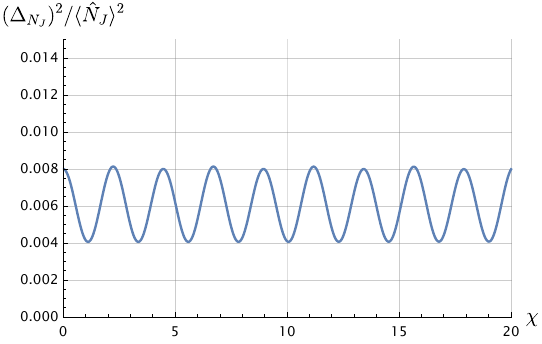}
\caption{Relative uncertainty of the number of particles in the coherent state representing the classically stationary state, for $\mathcal{K}^{(0)}_J = -1$, $\mathcal{K}^{(2)}_J = 1$ and $g=10^{-3}$. As expected for a coherent state, initially $(\Delta_{N_J})^2/\langle\hat{N}_J\rangle^2 = 1/\langle\hat{N}_J\rangle$; this initial value is actually an approximate upper bound for all times.}
\label{fig:pvar0}
\end{figure}

The mean-field approximation, which assumes that such a state remains coherent at all times, would imply that it is also stationary. This approximation is not exact, and since this is not an energy eigenstate we expect nontrivial time evolution. Nevertheless, for small $g$ the evolution of these states is almost stationary (see Figs.~\ref{fig:Nexp0}-\ref{fig:pvar0}). In particular, relative fluctuations stay very close to the initially small value, so that these states remain semiclassical under time evolution. In this sense, these quantum states behave classically enough to use them in the mean-field approximation. In terms of the physically relevant evolution of the volume, their properties are very similar to those of the exact ground state.

It is important to stress that this semiclassical behaviour of initially coherent states will not hold for arbitrary initial conditions. If we start with a coherent state with an expectation value of $\hat{\phi}_J$ far from the minimum of the potential, it evolves in a non-trivial way; the relative variance $\hat{N}_J$ (as well as of the fields $\hat{\varphi}_J$ and $\hat{\pi}_J$) increases, deviating from the classical behaviour. The mean-field approximation is then not applicable. We give an example of this in the Appendix. This is then the generic case in which the mean-field approximation breaks down in an interacting GFT, as previously discussed in Refs.~\cite{CosmoGFT,AxelGeneralCosmo}.

\section{Conclusions}

We have discussed several approaches for finding stationary cosmologies that could tentatively be associated with a Minkowski spacetime in the cosmological interpretation of GFT. First, we looked at a model in which the effective Newton's constant, related to the ``mass'' parameter $\mathcal{K}_J^{(0)}$ in GFT, is taken to zero. This theory could be seen as a potential starting point for a  standard model limit of GFT, in which matter propagates on an emergent spacetime but does not affect its structure, akin to setting $G = 0$ in the classical Einstein equations $G_{ab} = 8\pi G T_{ab}$. We saw that this model suffers from fine-tuning in the parameters, such that any small deviation turning the GFT dynamics into those of an inverted oscillator will develop instabilities generating an expanding Universe. We also found that the effective Friedmann equation does not imply an exactly stationary cosmology, but includes high-energy corrections similar to those responsible for a bounce in standard GFT cosmology.

We then studied a second approach which uses a GFT toy model including a quartic interaction term, equivalent to the dynamics of a double-well potential in usual quantum mechanics. By applying a numerical approximation technique in which the ladder operators and the Hamiltonian are represented as matrices, we found bound states starting from a ground state whose energy is very close to the classical minimum of the potential. We saw that such bound states are sharply peaked in the number of GFT quanta and hence in the volume, making them suitable candidates for semiclassical stationary cosmologies. In addition, the numerics show an approximately exponential relation of the relative variance of $\hat{N}$ and the coupling constant $g$: for theories with lower $g$ the ground state is even more sharply peaked around a given volume. Somewhat similar properties are found for the traditionally used Fock coherent states, set up in such a way that the expectation value of the group field sits in the classical minimum of the potential. These states evolve in time but only show small oscillations around the initial volume expectation value, with small fluctuations. While this nontrivial time evolution deviates from the mean-field approximation which would give exactly stationary expectation values, this deviation is small and the use of the mean-field approximation is justified. However, this only works for such a ``quantum gravity condensate'' in the minimum of the potential, and more generic initial conditions would lead to very non-semiclassical behaviour even for an initially coherent state.

The conclusion that bound states represent good candidates for a semiclassical cosmology in GFT is somewhat at odds with the traditional idea of using Fock coherent states for which uncertainties in the group field $\hat{\varphi}_J$ and momentum $\hat{\pi}_J$ can be made small. Since $\hat{\varphi}_J$ and $\hat{\pi}_J$ do not correspond to observables it seems more meaningful to demand that cosmologically relevant quantities such as the total volume or the energy (associated to a conjugate momentum of the matter scalar field), or more generally the $\mathfrak{su}(1,1)$ variables discussed in Ref.~\cite{AxelGeneralCosmo}, remain semiclassical. In this sense, our work suggests that more general classes of states may be considered to be viable candidates for GFT cosmology.

The numerical techniques applied here could be extended to more general and physically more interesting models for GFT cosmology, such as models with more general interaction terms or models coupling different Peter--Weyl modes. The only limitations come from computational cost, but since the calculations presented here were easy to implement there certainly seems to be scope for studying more involved cases.

{\em Acknowledgements.} --- This work was supported by the Royal Society through the University Research Fellowship UF160622 and University Research Fellowship Renewal URF$\backslash$R$\backslash$221005 (both SG). In addition, this work was supported in part by the Natural Sciences and Engineering Research Council of Canada and by Mitacs through the Mitacs Globalink Research Award (RS).

\appendix
\section{Coherent states with generic initial conditions}
\begin{figure}[htbp]
\centering
\includegraphics[width=1\columnwidth]{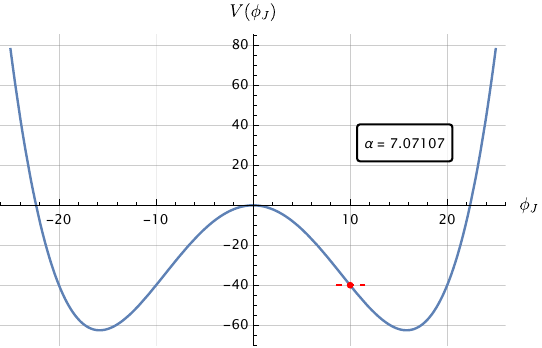}
\caption{Classical potential and initial expectation value of the field (red point). We set the parameters of the Hamiltonian to $\mathcal{K}^{(0)}_J = -1$, $\mathcal{K}^{(2)}_J = 1$ and $g=10^{-3}$.}
\label{fig:unstabpot}
\end{figure}

Here we consider the previous interacting case from Eq.~\eqref{interactHam} and initial coherent state $\ket{\alpha}$ associated with an expectation value of the field $\bra{\alpha}\hat{\varphi}_J\ket{\alpha}$ located far from the classical minimum of the potential (see Fig.~\ref{fig:unstabpot}). We choose $\alpha=10/\sqrt{2}\sim 7.07$ as an example for an initial field value away from the minimum of the potential, but the behaviour we observe here appears to be generic.

The quantum evolution is given by the Schr\"odinger equation in the truncation described around Eq.~\eqref{eq:basis}. With a dimension of $900$ we can calculate the numerical solution for the aforementioned initial condition with sufficient speed and precision.

\begin{figure}[htbp]
\centering
\includegraphics[width=1\columnwidth]{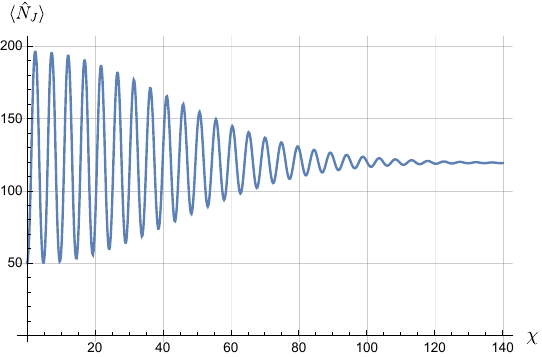}
\caption{Evolution of the particle number in the unstable case. Due to quantum effects of the quartic potential, the evolution results in a damping of the initial oscillations.}
\label{fig:noscexp}
\end{figure}

\begin{figure}[htbp]
\centering
\includegraphics[width=1\columnwidth]{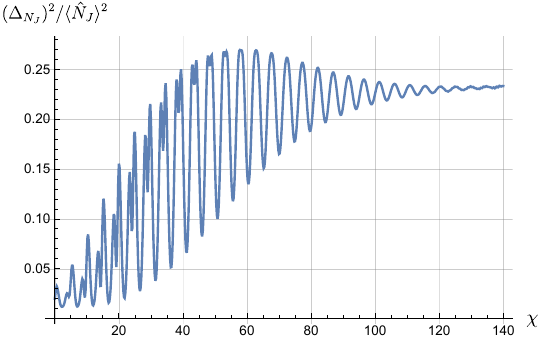}
\caption{The relative variance of the particle number increases in time and seems to converge to a large value, meaning that the state is not close to a coherent state and the semiclassical (or mean-field) interpretation is lost.}
\label{fig:noscvar}
\end{figure}

Classically, we would expect the field to oscillate between $\phi_J=10$ and $\phi_J=20$, corresponding to particle numbers of 50 to 200. In Fig.~\ref{fig:noscexp} we see that this is indeed what happens initially, but after a short time the behaviour of the particle number (and therefore the volume) differs from this classical expectation: the oscillations become damped leading to an asymptotic value around the minimum of the potential. At the same time, the state no longer remains peaked in the volume, and acquires large fluctuations (Fig.~\ref{fig:noscvar}). These results demonstrate the breakdown of the mean-field approximation for this kind of states, unlike what we found for states peaked initially at the minimum of the potential.

\end{document}